\documentclass[aip,jcp,amsmath,eqsecnum,unsortedaddress,nofootinbib,floatfix,10pt,preprint]{revtex4-1}
\usepackage{amsmath}
\usepackage{graphicx} 
\usepackage{dcolumn}%
\usepackage{bm}%
\usepackage{color}

\begin{document}

\title{Solvent Effects in the Helix-Coil Transition Model can explain the Unusual Biophysics of Intrinsically Disordered Proteins.}

\author{Artem Badasyan}
\email{abadasyan@gmail.com}
\affiliation{Materials Research Laboratory, University of Nova Gorica,\\ Vipavska 13, SI-5000 Nova Gorica, Slovenia, EU}

\author{Yevgeni Sh. Mamasakhlisov}
\affiliation{Department of Molecular Physics, Yerevan State University,\\ A. Manougian Str.1, 0025, Yerevan, Armenia}

\author{Rudolf Podgornik}
\affiliation{Department of Theoretical Physics, J. Stefan Institute and Department of Physics, Faculty of Mathematics and Physics, University of Ljubljana - SI-1000 Ljubljana, Slovenia, EU and Department of Physics, University of Massachusetts, Amherst, MA 01003-9337 USA}

\author{V. Adrian Parsegian}
\affiliation{Department of Physics, University of Massachusetts, Amherst, MA 01003-9337 USA}

\date{\today}

\begin{abstract}
We analyze a model statistical description of the polypeptide chain helix-coil transition, where we take into account the specificity of its primary sequence, as quantified by the phase space volume ratio of the number of all accessible states to the number corresponding to a helical conformation. The resulting transition phase diagram is then juxtaposed with the unusual behavior of the secondary structures in Intrinsically Disordered Proteins (IDPs) and a number of similarities are observed, even if the protein folding is a more complex transition than the helix-coil transition. In fact, the deficit in bulky and hydrophobic amino acids observed in IDPs, translated into larger values of phase space volume, allows us to locate the region in parameter space of the helix-coil transition that would correspond to the secondary structure transformations that are intrinsic to conformational transitions in IDPs and that is characterized by a modified phase diagram when compared to globular proteins. Here we argue how the nature of this modified phase diagram, obtained from a model of the helix-coil transition in a solvent, would illuminate the {\sl turned-out} response of IDPs to the changes in the environment conditions that follow straightforwardly from the re-entrant (cold denaturation) branch in their folding phase diagram.
\end{abstract}

\maketitle

\section{Introduction}

An extensive revision of ideas and paradigms underpinning the proper understanding of protein folding is underway \cite{idp14,tompa}. Arguably, among the other important facts, illustrating the apparent lack of understanding of protein folding, is the existence of Intrinsically Disordered Proteins (IDP) in a range of environmental conditions, where 'regular' globular proteins have a well defined and weakly fluctuating secondary and tertiary structural elements. Clearly  since IDPs are persistent, there must be some evolutionary advantage of proteins being disordered \cite{zhirong}. On the other hand, structurally IDPs belong to the same family of natural polypeptides as other proteins and it would appear reasonable that they should behave similarly. But do they? 

In order to address this and related questions we will take recourse in the {\sl Generalized Model of Polypeptide Chain} (GMPC) \cite{biopoly1,biopoly2} of the helix-coil transition, that will be taken as a proxy for the protein folding transition, and show that the salient features of the behavior of IDPs can be described by different values of a single entropic parameter, germane to the model. This parameter quantifies the degeneracy of the coil conformation and on this level of description, the difference between regular proteins and IDPs  boils down to the differences in the ratio of the number of all accessible states and the number of states actually available for the repeating unit in a helical conformation, which we show to be increased in the case of IDPs. While our argumentation will be based on this phenomenological approach it will allow us to rationalize in a straightforward way many of the properties of IDPs that would be difficult to rationalize in more detailed but also more approximate fully microscopic models.

Since IDPs are unable to fold autonomously into specific structures, their structural description requires conformational ensemble-based approaches and leaves little workspace for topology-oriented Go-like modeling of folding. Most of experimental studies of IDPs utilize spectrophotometric techniques, which record the changes in secondary structure content. Secondary structure formation in proteins in general takes place simultaneously with the formation of tertiary structure, making it clear that these two events are intimately related. We intentionally leave aside the discussion of kinetic mechanism of folding (on what is formed first: secondary or tertiary structure), which is an interesting question, but is beyond the scope of our study. The important feature we reiterate is that both levels of structural organization are coupled and changes in secondary structure usually happen under the same external conditions, where the changes in the tertiary structure take place. Therefore the explanation of peculiarities of secondary structure formation in IDPs and globular proteins in general, is not only relevant per se, but would also shed light on relevant differences of phase diagrams of folding of these systems.

\begin{figure}[t!]
	\begin{center}
	\includegraphics[width=\columnwidth]{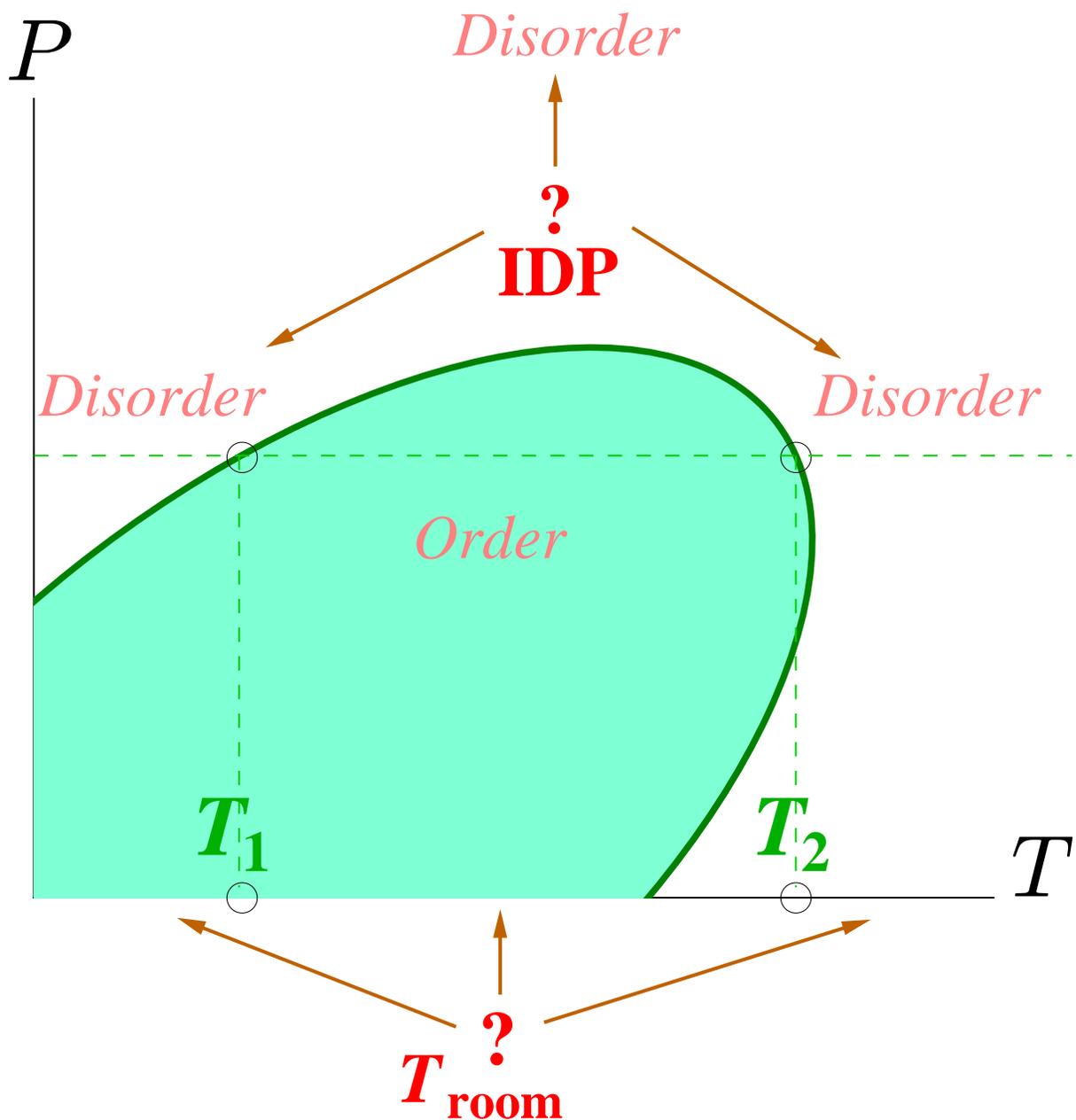}
	\caption{(Color online) Schematic representation of the "skewed ellipsoid" $P-T$ cross section of the protein folding phase diagram, after Hawley \cite{hawley}. The arrows indicate possible positioning of the IDPs in this phase diagram.}
	\label{f0}
	\end{center}
\end{figure}

The effect of environmental conditions on the structural states of proteins is most naturally described {\sl via} the induced changes in {the phase diagram}. Depending on the particular experimental set-up and relevant experimental variables, different cross sections through the phase diagram can be studied, among them most notably the pressure-temperature and the osmotic pressure-temperature dependencies. As is well known \cite{hawley,smeller}, pressure-temperature phase {diagrams of proteins} can be roughly approximated by a "skewed ellipsoid", see Fig. \ref{f0}. At a fixed pressure (usually ambient pressure) the shape of the experimental phase diagram reveals two distinct temperature regions with two transition points: {\sl cold denaturation} at low temperatures and {\sl unfolding} at high temperatures. Recent single-molecule F\" orster resonance energy transfer (FRET) experiments with heat-unfolded states of yeast frataxin, HIV integrase, Csp M34, $\lambda$-repressor, and two prothymosin fragments, provide additional evidence of the existence of both cold and heat denaturation as an intrinsic feature of polypeptides \cite{aznauryan,wuttke}.  

In his review Uversky \cite{uver1} argues that IDPs have a "turned-out" response to changes in their environment, in the sense that under environmental conditions which result in unfolding of regular globular proteins, IDPs actually show a tendency toward more structure, as clearly demonstrated in the case of $\alpha$-synuclein, caldesmon 636-771 fragment, phosphodiesterase $\gamma$-subunit, the receptor extracellular domain of nerve growth factor, $\alpha$-casein, and several other IDPs (see Refs. \cite{uver1,uver2} and references therein). Experimentally \cite{uver1,uver2} it seems that the behavior of IDPs is consistent with the low-temperature region of the phase diagram as some IDPs gain structure and aggregate with increasing temperature, a phenomenon clearly related to cold denaturation \cite{sanfelice}.  

Another important example of the unusual response to environmental conditions, still lacking a proper explanation, is the absence of conformational changes under external osmotic stress in some IDPs, as summarized in a recent review on IDPs' response to crowding \cite{kart}.

On the one hand, it thus appears that the behavior of ordinary globular proteins and IDPs differ fundamentally in terms of protein conformations at set values of environmental conditions. On the other hand, independently of whether the particular protein belongs to a loosely defined 'globular' or 'intrinsically disordered' class, one would expect a qualitative similarity of behavior among all the proteins since they do belong to the same structural class of polypeptides, differing only in the particular amino-acid residue sequence (primary structure). This similarity should be clearly distinguishable also in the very shape of the phase diagram of folding, determined solely by the protein chemical structure and interactions that govern the folding event. Nevertheless, the experiments show different behaviors for globular proteins and IDPs in the same range of environmental conditions. 

For the schematic phase diagram in Fig. \ref{f0} to agree with these experimental results, one should then have the temperature of cold denaturation to be around room temperature ($T_1 \approx T_{room}$), which would be quite surprising as most globular proteins in fact have unfolding happening around room temperature ($T_2 \approx T_{room}$), and thus exhibit only the second, direct order-disorder transition within the physiological range. In order to elaborate the detailed positioning of the IDPs in the elliptic phase diagram, we now need to invoke a specific model, describing the secondary structure formation, that would be sufficiently rich to allow the implementation of a re-entrant phase diagram and could account on some level also for the peculiarities of the primary sequence of IDPs.

\section{Model}

The studies of helix-coil transition in polypeptides have started more than sixty years ago. It was not realized then and became apparent only later, that there is a crucial difference between the helix-coil transition and the protein folding \cite{gros}. Since the helical structures are mainly stabilized by hydrogen bonds (H-bonds), which are of short range in space and are formed between near neighboring repeat-units of a linear polypeptide molecule, the Landau-Pierls theorem holds true and no phase transition results. For this reason the helix-coil transition takes place in a finite interval of external parameter values (\emph{e.g.} temperature) \cite{bad10}. Instead, protein folding can be considered as an example of a coil-globule transition, which for rigid chains is a first-order phase transition  \cite{gros}. In practice, since the natural proteins are of finite size and structurally disordered (heterogeneous), folding transition also takes place in a finite interval. The principal difference between the two types of transition (helix-coil \emph{vs.} coil-globule) is that with increasing chain length (quantified by the number of repeating units $N$), the interval of the helix-coil transition tends to be finite, while the interval of the folding transition should tend to vanish. Nevertheless, although different in nature, these two types of transition are tightly interconnected. It follows from general polymer consideration that the coil-globule transition essentially depends on the rigidity of the chain, and the formation (destruction) of a helix in a region of a polypeptide chain significantly increases (decreases) its local rigidity. How exactly these two levels of protein structural organization are related, is not fully understood yet. What is experimentally known is that both secondary and tertiary structure transformations take place in a coordinated manner, in the same interval of external conditions.

The initial paradigm of protein science, as proposed by Pauling and Corey in the early 1950s \cite{pauling}, claimed that the conformations of biopolymers are controlled by intra-chain polypeptide H-bonding stabilizing the structural elements of proteins, a point of view clearly endorsed also in the seminal papers of Zimm and Bragg \cite{zb}, Lifson and Roig \cite{lr} as well as others \cite{polsher}. Later, Kauzmann \cite{kauzmann} generalized this paradigm by widening the range of interactions that stabilize the configurations of polypeptide chains from purely intra-chain H-bonding to polypeptide-water H-bonding as well, since disproportionately strong H-bonds in water themselves give rise to {\sl hydrophobic interactions}. This "hydrophobic effect" came to  dominate the thinking about protein folding until quite recently, when a more balanced view of different types of H-bonding reconciled the contribution of intra-peptide and water-polypeptide H-bonding \cite{dill,bennaim,dias}.

Consequently some models, like the Hydrophobic-Polar (HP) model \cite{dill1}, were designed specifically to account for the hydrophobic interaction, while others, like the Zimm-Bragg model \cite{zb}, set out to implement only intra-chain H-bonding, with many other attempts specifically aimed at a proper description of the balance of both types of H-bonding, the intra-chain as well as the polypeptide-water hydrogen bonds, seen as simultaneously governing the protein folding event \cite{dill,bennaim,dias} through secondary structure formation and side chain packing \cite{dill2}. With this realization of the fundamental importance of H-bonding in protein folding, the propensity of water to make hydrogen bonds was singled out by Dill \cite{dill} also as the determining factor giving rise to the overall "skewed ellipsoid" phase diagram of protein folding \cite{hawley}. It thus seems appropriate to focus on the fundamental role of H-bonding also in the secondary structure elements of IDPs \cite{uver1,uver2}. 

In what follows we will use a detailed analysis of a certain implementation of the helix-coil transition as a proxy to actually study some aspects of the folding transition. While protein folding is certainly a far more complex transition than the helix-coil transition, it is interesting to explore the behavior within  the helix-coil paradigm that matches some aspects of that exhibited by IDPs in folding.

\subsection*{Solvent-free model describing H-bonding in polypeptides}

The {\sl Generalized Model of Polypeptide Chain} (GMPC) \cite{biopoly1,biopoly2} of the helix-coil transition was recently shown to possess a sufficiently rich phase diagram structure that can exhibit re-entrant behavior as a general feature of any polypeptide chain, controlled by the competition between intra- (within the polypeptide) and inter-molecular (polypeptide-water) H-bonding \cite{bad11,bad14}. Details of this  spin-based model as well as the relations with other models of folding were elaborated previously \cite{biopoly1,biopoly2,bad10,EPJE}.  The re-entrant behavior exhibited by GMPC is one of its general features and should be present for both globular proteins as well as IDPs, since they both belong to the same structural class of natural polypeptides. A yet unresolved issue is then to locate the parameter region in the schematic phase diagram of folding that corresponds to the IDPs and, based on that, to explain how the peculiarities of the primary sequences of IDPs and globular proteins lead to differences in their behavior at normal conditions (room temperature and ambient pressure).

Statistical description of the helix-coil transition within  spin-based models requires three parameters: the energy parameter $W=\exp(U/T) = e^{J}$, where $U$ is the energy of the hydrogen bond and $T$ is temperature; a geometric parameter $\Delta$ that describes the geometry of hydrogen bond formation, usually involving three successive repeat-units; and an entropy parameter $Q$, that quantifies the degeneracy of coil conformation. The type of amino acid enters directly only via the entropic $Q$ parameter, and it is therefore appropriate to consider it in greater detail. 

Since for polypeptides the bond lengths and bond angles do not vary much and the peptide bonds are planar, the conformation of amino acids in the chain can be described by only two torsional angles $(\phi,\psi)$ \cite{flory}. The corresponding energy landscape of the different conformations then follows from the $(\phi,\psi)$ Ramachandran plots \cite{rama}. In this framework one can relate an area of the $(\phi,\psi)$ phase space corresponding to the alpha-helical conformation with the partition function of a single amino acid corresponding to such conformation. It then follows that the entropic $Q$ parameter is given by the ratio of the number of all accessible states and the number of states actually available for the repeating unit in a helical conformation \cite{biopoly1}
\begin{equation}
\label{Qdef}
Q(T)=\frac{\int_{\Gamma}e^{-E(\phi,\psi)/T}d\phi d\psi}{\int_{\Gamma_\alpha}e^{-E(\phi,\psi)/T}d\phi d\psi},
\end{equation}
{\sl i.e.}, as the ratio of the full partition function for a particular amino acid, normalized by the partition function for the same amino acid in the alpha-helical region of the Ramachandran plot. In principle, Eq.~\ref{Qdef} suggests a general way to calculate the values of parameter $Q$ for any primary structure.
If one chooses a hard-sphere repulsive interaction as the simplest possible interaction potential in Eq.~\ref{Qdef}, this yields a temperature-independent entropic parameter
\begin{equation}
\label{Qhs}
Q=\frac{\Gamma}{\Gamma_\alpha},
\end{equation}
equal to the ratio between the available areas in the $(\phi,\psi)$ phase space obtainable from the Ramachandran plots. For more complicated, yet more realistic and less well characterized interaction potentials, the value of $Q$, while still defined in the same way, would be that much more difficult to calculate and requires a separate thorough study. In the absence of such study one should see it as a {\em phenomenological parameter} whose value can be deduced from a comparison with experiments.

On the analytical level the partition function $Z_{0}(W,Q)$ of the GMPC is obtained via the {\sl transfer-matrix formalism} that yields a set of eigenvalues $\{\lambda_i\}$ (${i=\overline{1,\Delta}}$) obtained from the characteristic equation (for details see Ref.~\cite{biopoly1,EPJE})
\begin{equation}
\label{chareq}
\lambda^{\Delta-1}(\lambda-W)(\lambda-Q)=(W-1)(Q-1),
\end{equation}
\noindent so that 
\begin{equation}
\label{partfuncbasictm}
Z_{0}(W,Q) = \text{Trace } \hat{G}^N=\sum\limits_{k=1}^{\Delta}\lambda_{k}^{N}\simeq\lambda_{1}^{N},
\end{equation}
\noindent where $\hat G(\Delta \times \Delta)$ is the transfer matrix of the model. It immediately follows that
\begin{equation}
\label{fenergy}
F_0(W,Q) \simeq -NT\log\lambda_{1} = - NT\log\max\{W,Q\},
\end{equation}
\noindent since $W,Q$ play the role of asymptotes for the two largest eigenvalues. The form of Eq.~\ref{chareq} entails that changes in both the entropic parameter $Q$ as well as the energetic parameter $W$ affect the equilibrium properties in exactly the same way. Thus in principle, the same effect on a polypeptide chain conformations can be achieved by changing either one of them. 
\begin{figure}[!ht]
\begin{center}
\includegraphics[width=\columnwidth]{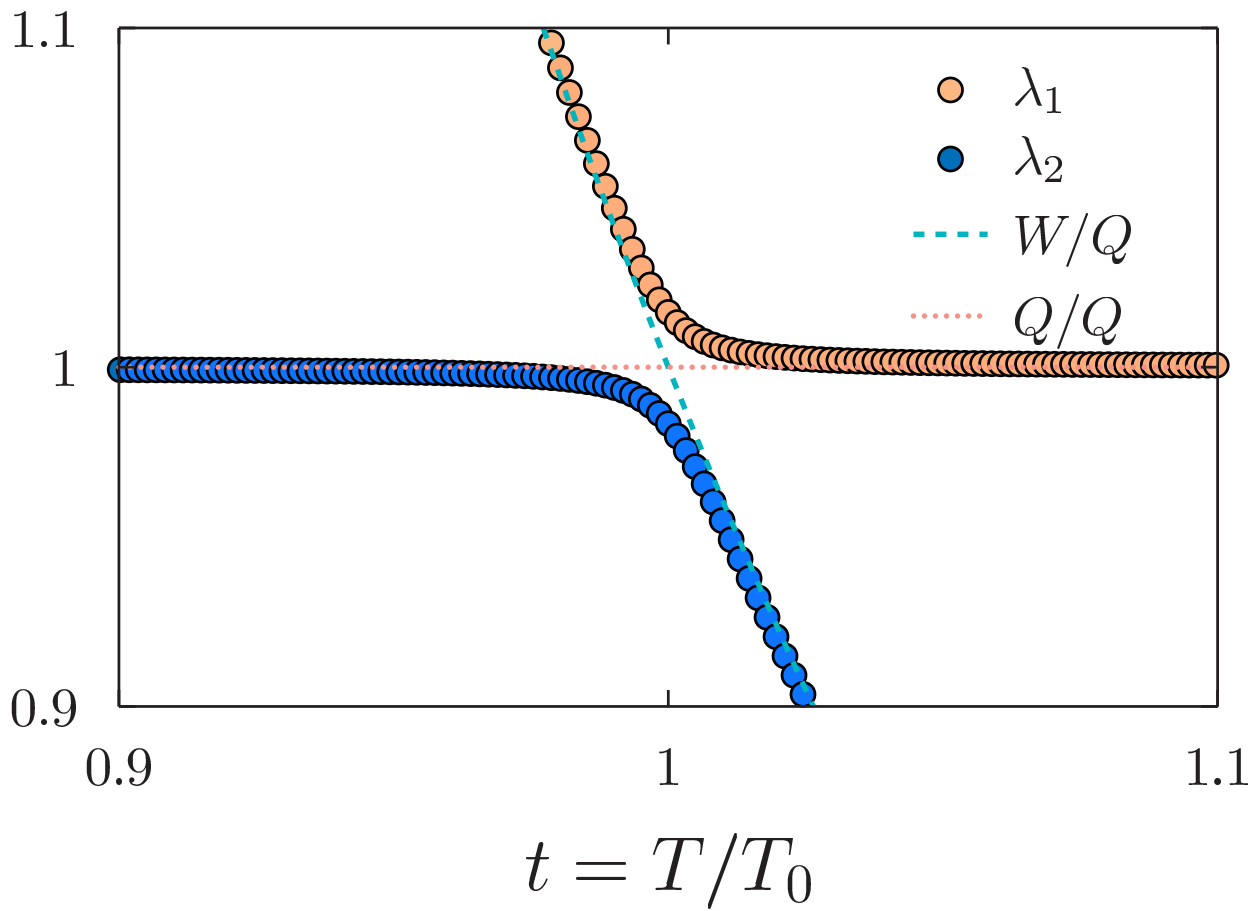}
\caption{(Color online) The temperature dependence of the largest ($\lambda_1$) and second largest ($\lambda_2$) eigenvalues obtained from Eq.~\ref{chareq}, energetic parameter $W=\exp{(1/t)}$ and entropic parameter $Q$ in dimensionless units. Shown is the solvent-free, \emph{in vacuo} case with $\Delta=3,Q=60$.  The obvious asymptotes for $\lambda_1$ are $\max\{W,Q\}$ and for $\lambda_2$ they are $\min\{W,Q\}$.}
\label{fchareq} 
\end{center}
\end{figure}
In the thermodynamic limit it suffices to study the temperature dependence of the two largest eigenvalues that determine the free energy and the correlation length of the system, and furthermore define the stability and the cooperativity of the helix-coil transition \cite{EPJE}. The eigenvalues approach closest together at a point where the asymptotes $W(T)$ and $Q$ cross (Fig.~\ref{fchareq}), in accordance with general physical considerations that the transition takes place when the entropy and the energy compensate each other. 

\subsection*{Changes in the model due to the solvent}

Hydrogen bond formation in polypeptides is known to affect three successive repeat-units so that $\Delta=3$ in {\sl any} solvent. On the contrary, the energetic parameter $W$ and the entropic parameter $Q$, which correspond to the number of microstates involved in the coil conformation, {\sl i.e.}, the degeneracy of the coil, can be affected by the presence of the solvent. It can be shown \cite{biopoly1,bad11,PRL,bad14} that the effect of the solvent can be accounted for within the exact same formalism, as detailed above, simply by re-normalizing the model parameters, $W \longrightarrow \widetilde{W}$ and $Q \longrightarrow \widetilde{Q}$, where
\begin{equation}
\label{Wrenorm}
\widetilde{W} = e^{U_{eff}/t}= \frac{q^{2}e^J}{(q+e^{I}-1)^{2}}=
\left( \frac{q \, e^{1/2t}}{q+e^{\frac{1+\alpha}{2t}}-1}\right ) ^{2} 
\end{equation}
\noindent and
\begin{equation}
\label{Qrenorm}
\widetilde{Q}=1+(Q-1) \frac{p+e^{I_c}-1}{p+e^{I_h}-1}=1+(Q-1) \frac{p+e^{\alpha_c/t}-1}{p+e^{\alpha_h/t}-1}.
\end{equation}
Above $t=2T/(U_{pp}+U_{ss})$, $I_{c,h} = E_{c,h}/T$, $\alpha=\frac{2U_{ps}-(U_{pp}+U_{ss})}{U_{pp}+U_{ss}}$, and $\alpha_{h,c}=\frac{2E_{h,c}}{U_{pp}+U_{ss}}$, where $U_{pp,ss,ps}$ are the energies of intra- (polymer-polymer) and inter-molecular (solvent-solvent and polymer-solvent) H-bonds, respectively, while $E_{h,c}$ are the changes in the helix and coil energy of repeat unit due to the external osmotic stress \cite{PRL}. 

Naturally, the renormalized parameters $\widetilde{W}$ and $\widetilde{Q}$ continue to play a role of asymptotes for the two largest eigenvalues that determine the behavior of the model, but with a much different temperature dependence compared with the solvent free case, see Fig. \ref{fchareq} (also consult \cite{bad14} for a more detailed discussion). This re-normalization then redefines the transfer matrix, the partition function Eq.~(\ref{partfuncbasictm}) and the characteristic equation Eq.~(\ref{chareq}), accounting fully for the solvent effects within the GMPC model. The most important consequence of this solvent re-normalization of the GMPC model parameters is that the two largest eigenvalues approach closely not only at one, but at {\sl two temperatures}, the second one corresponding to a signature of the {\sl cold denaturation}. Varying $\alpha$ and $\Delta \alpha=\alpha_{h}-\alpha_{c}$ then leads to a sequence of conformational transitions characterized by very complicated phase diagrams \cite{bad14}. In fact, changes in $\alpha$ shift the balance between the inter- and intra-molecular hydrogen-bonds and can thus be rationalized as mimicking pressure effects \cite{bad11}, while changes in $\Delta \alpha$ can be rationalized as describing osmotic stress effects\cite{PRL}. 

\begin{figure}[!ht]
\begin{center}
\includegraphics[width=\columnwidth]{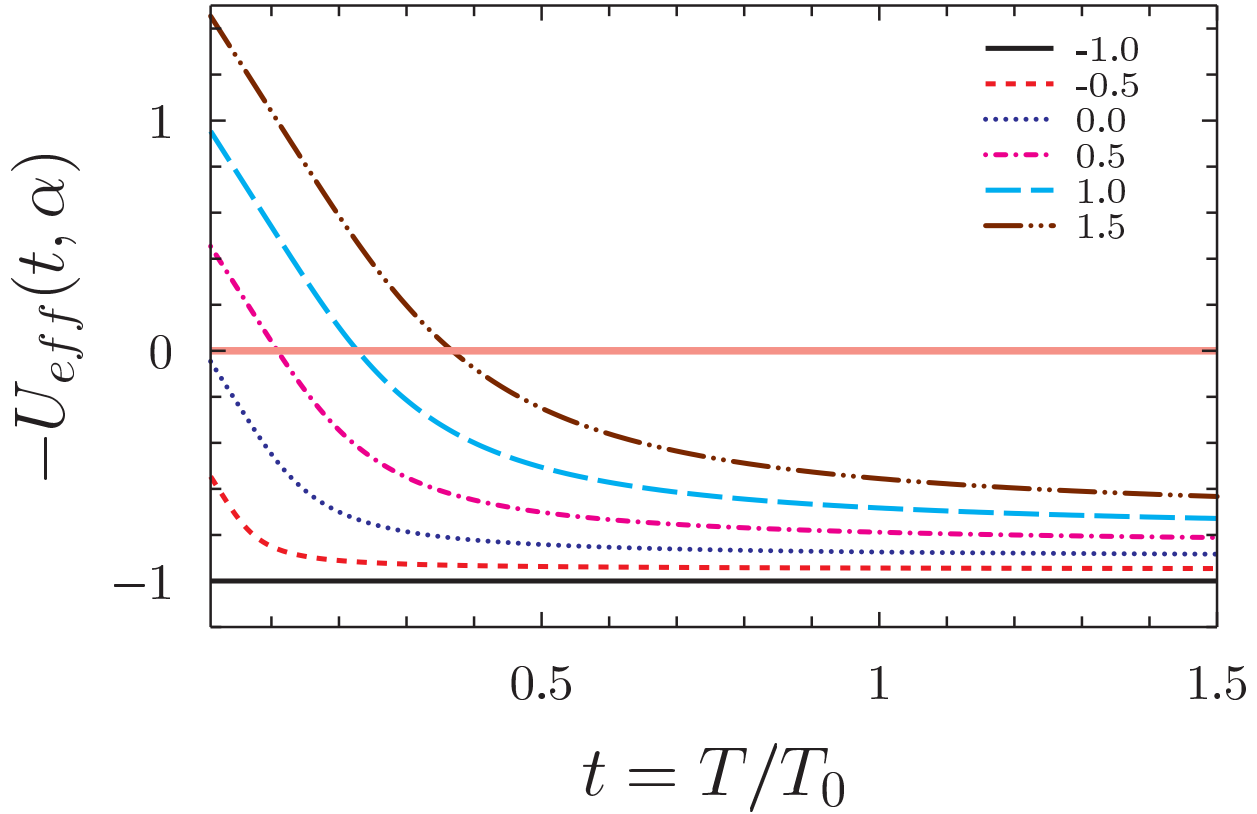}
\caption{(Color online) The temperature dependence of the "effective energy" of H-bonding, $-U_{eff}$. Numbers shown correspond to different values of parameter $\alpha$. Obviously even the sign of effective hydrogen-bond interaction can change as a function of temperature.} 
\label{ueff}
\end{center}
\end{figure}

The re-normalization of the temperature-reduced energy of H-bonding between polymeric units from the solvent-free value $J$ to the solvent value is characterized by $J \longrightarrow \widetilde{J} = U_{eff}(\alpha,t)/t$ \cite{bad14}. In this context it is informative to investigate the dependence of the "effective energy" of the hydrogen-bond formation, $-U_{eff}$ from Eq.~(\ref{Wrenorm}), on the temperature at different values of $\alpha$. As shown in Fig.\ref{ueff}, $U_{eff}$ can even change sign with increasing temperature, implying that the balance between the intra- and the inter-molecular interactions changes with the temperature through conversion of the intra-protein H-bonds to H-bonds with water, which in its turn explains also the appearance of cold denaturation in the phase diagram. 
\begin{figure*}[!ht]
	\begin{center}
		\begin{minipage}[b]{0.48\textwidth}
			\begin{center}
\includegraphics[width=\textwidth]{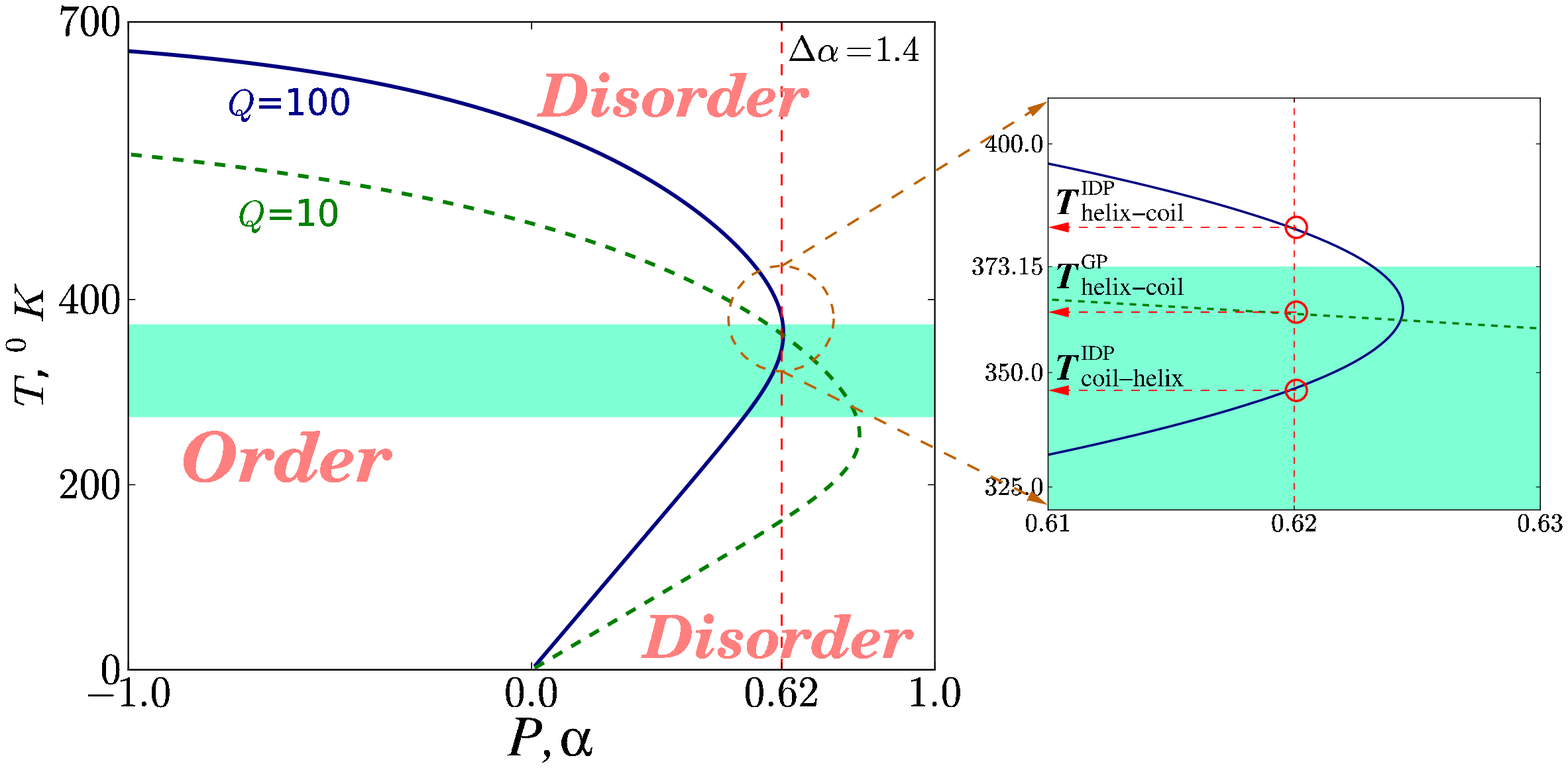} (a)
			\end{center}
		\end{minipage}
		\begin{minipage}[b]{0.48\textwidth}
			\begin{center}
\includegraphics[width=\textwidth]{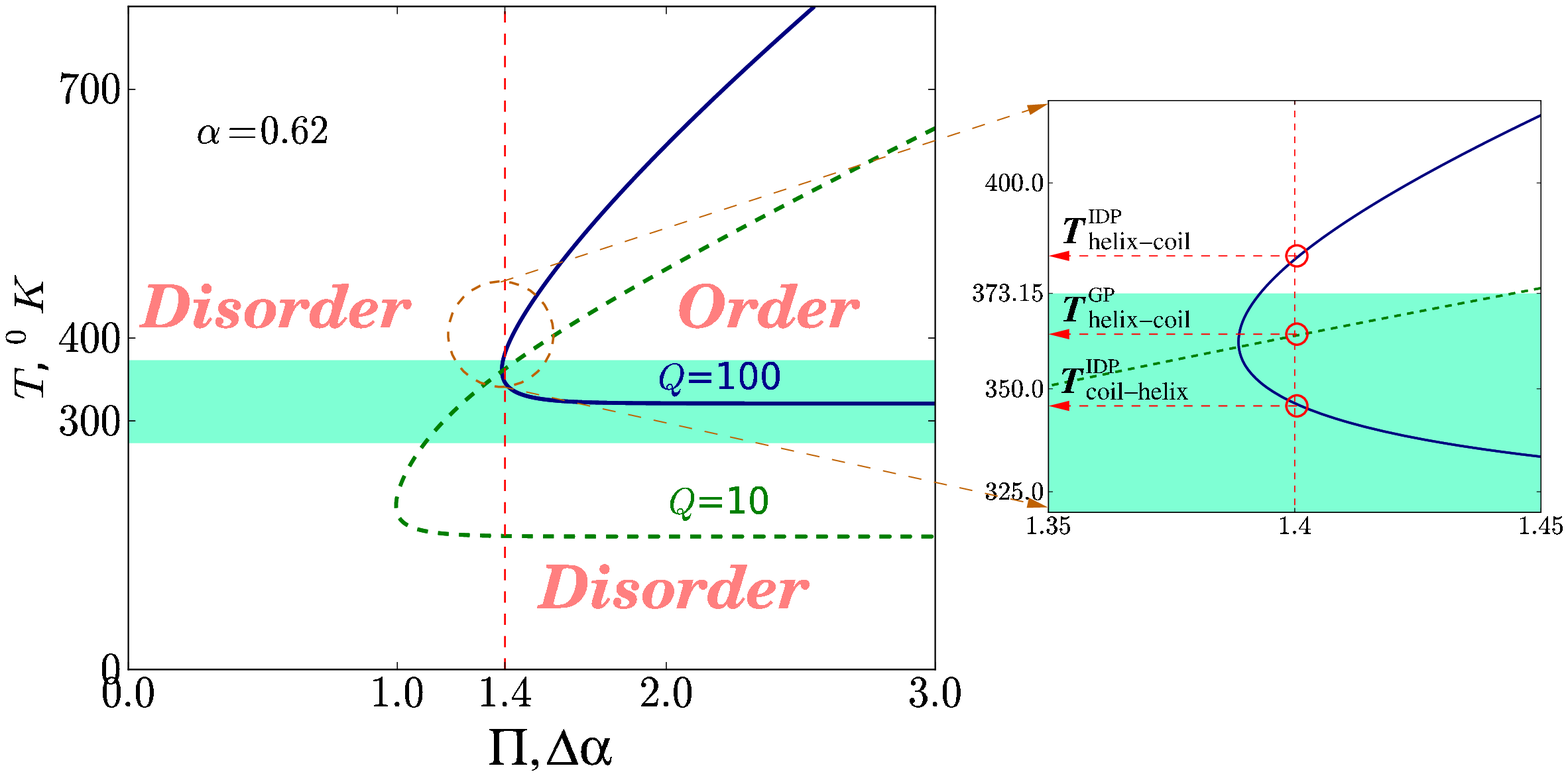} (b)
			\end{center}
		\end{minipage}
\caption{(Color online) Phase diagrams obtained by numerically solving $\widetilde{W}=\widetilde{Q}$ equation using Sage \cite{sage}. Temperature is in degrees Kelvin, pressure in dimensionless units of $\alpha$ and osmotic stress in dimensionless units of $\Delta \alpha$. The following set of parameters is used: $\Delta=3$, $q=10$, $p=100$. Insets show the presence of the helix-coil transition for globular protein ($Q=10$) and coil-helix or cold denaturation for IDP ($Q=100$) within the range of physiological temperatures. Outside of this physiological range, above the water boiling point, is the helix-coil transition for IDP, and below the water freezing point the cold denaturation of globular proteins (not shown). The projection of phase diagram (a) in the pressure - temperature plane is plotted at $\Delta \alpha=1.4$ and (b) in the osmotic stress - temperature plane is plotted at $\alpha=0.62$.}
\label{ff}
	\end{center}
\end{figure*}

\section{Results and Discussion}

Before we continue with a detailed discussion of the above results, we find it relevant to reiterate that they resulted from an implementation of the GMPC model for helix-coil transition with the account of water-polymer interactions and osmotic stress, as a proxy to study the more complicated protein-folding transition. While the differences between these two transitions are important and were elucidated above, we nevertheless find it interesting that the GMPC model of the helix-coil transition exhibits features that also capture some of the relevant differences in the peculiarities of the phase diagrams for folding transitions of IDPs. Moreover, since most of the experimental studies on IDPs have been performed using the spectrophotometric methods \cite{uver1,uver2}, which mainly capture the secondary structure transformations, the application of a theory that describes the secondary structure formation seems at least reasonable if not wholly appropriate.

\subsection*{IDPs and globular proteins differ in terms of $Q$ only}

First, {the} general properties of the GMPC model can illuminate the difference between IDPs and globular proteins. From the very definition of IDPs, it is clear that they are optimized to maintain disorder even under environmental conditions where the 'regular' proteins would be ordered, {\sl i.e.}, folded. Indeed, the only difference between both types of proteins is the relative abundance of different types of amino acids in their primary sequence, with deficit of bulky amino-acids and more abundant flexible, disorder-prone amino acids in IDPs \cite{dunker}. 

Within the GMPC model the type of amino acid enters directly only {\sl via} the entropic $Q$ parameter. The available areas on the Ramachandran plot for different amino acids vary depending on how bulky the amino acids are, with a general tendency of decreasing for bulkier amino acids as compared to light ones, purely due to steric exclusion. Furthermore, since homo-polymeric polypeptides are rare, the quenched compositional disorder (primary structure) must also be accounted for by assigning some average value of $Q$ to the polypeptide under study. 

While exact calculation of $Q$ from microscopic parameter is thus unrealistic at this stage, it is nevertheless clear on the qualitative level that the presence of less bulky amino-acids in the primary sequence of IDPs will eventually lead to an increase of $Q$ values as compared with globular proteins with higher content of massive amino acids. From the fundamental definition of entropy and according to Eq.~\ref{Qhs}, one can furthermore think of $\log (Q)= \log \Gamma - \log \Gamma_\alpha$ as $\Delta S=S_{coil}-S_{helix}$, {\sl i.e.} the difference between the entropy of the coil state and the helical state, and consider higher values of $\Delta S$ for sequences with less bulky amino acids. This leads to a general qualitative conclusion that one should have $Q_{IDP} > Q_{Glob.Prot.}$. For the sake of comparison we assign $Q_{IDP}=100$ and $Q_{Glob.Prot.}=10$, both values well within the relevant range. The relation to the Zimm-Bragg cooperativity parameter can be established as $\sigma(Q=10)\approx 2.5\times 10^{-3}$ and $\sigma(Q=100)\approx 2.5\times 10^{-5}$, within the experimentally verified reasonable range \cite{biopoly1,biopoly2}.

\subsection*{Protein reply to the solvent and osmotic effects depends on $Q$}

Assembling all the results together, the GMPC model of the helix-coil transition sheds some light on the shape of the folding phase diagram of IDPs and its relation to that of the "globular" proteins. Comparison of parts of the "skewed ellipsoid" pressure-temperature and osmotic pressure - temperature cross sections of the phase diagrams of IDPs in Figs. \ref{ff} for different values of the entropic parameter $Q$ shows that while the overall shapes of the phase diagrams are the same as expected, the temperatures of both the direct helix-coil and the re-entrant coil-helix transitions (at the same $\alpha$ or $\Delta \alpha$ values) are shifted to higher values in the case of IDPs. This brings about a counter intuitive situation that IDPs can be stabilized at environmental conditions, which would appear as extremely destabilizing for 'globular' proteins, a clear example of the {\sl s.c.} "turned out response" of IDPs \cite{uver1}. Moreover, there exists a range of pressure values(in terms of parameter $\alpha$, describing the balance between intra- and inter-molecular hydrogen-bonding) where regular "globular" proteins still allow for a coil-helix-coil transitions, but IDPs remain within a disordered, coil state at any temperature. The effect of applied osmotic stress for the two types of proteins also differs markedly. For any polypeptide in water (with $\alpha > -1 $), there exists a minimal value of osmotic stress ($\Delta \alpha$, in the physical units), below which no ordered state can exist. This could in fact be one of the main reasons for controversy in experimental observations of osmotic stress effects for IDPs \cite{kart}. For "globular" proteins this value is much smaller than for IDPs, so that there must exist an osmotic stress window where globular proteins can be observed in an ordered state, while IDPs will be disordered. Same happens in the case of temperature: there exists a minimal value of temperature, below which no order exists (due to destabilizing role of water), which is higher for IDPs. Then it seems that the proposal \cite{uver2} to classify the proteins by their response to osmotic stress, namely, to consider the absence of stabilization as a sign of IDP-like behavior, is ill-defined, since it would depend on the temperature at which every particular experiment was conducted.

Summarizing the above, we can state that within the physiologically relevant range of environmental parameters, the disordered regime is more common for polypeptide chains with higher flexibility (large $Q$, IDPs), than for more rigid chains (smaller $Q$, globular proteins). This could well be the experimental region of pressures/temperatures where Tompa {\sl et al}. \cite{tompa} have reported the absence of a globular state for the IDP that they have studied and came to the conclusion that the disordered state must be physiologically functional. Furthermore, Binolfi, Theillet and Selenko \cite{selenko} using bacterial in-cell NMR have reported that $\alpha$-synuclein is disordered in the cell \emph{in vivo}, and Soranno {\sl et al}.\cite{soranno}, have shown that synthetic polymeric crowders of high molecular weight cannot compact highly charged IDPs beyond a certain threshold and that molecular crowding does not affect all IDPs equally. Our Fig.~\ref{ff} b) allows to explain the absence of the globular state for some IDPs and suggests that to achieve the compaction of an IDP by crowder, it can be sometimes useful to perform the experiment not at ambient, but at a higher value of temperature.

\subsection*{Relationship between $Q$ and other compositional parameters}
In Ref.~\cite{idp14} a diagram-of-states was suggested based on several compositional parameters: 'hydropathy' or 'hydrophobicity' and the fraction of positively and negatively charged residues (for updated review see also \cite{pappu}). While the presence of charge and therefore the polyampholyte/polyelectrolyte nature of IDPs are no doubt important, we would like to stress that amino-acid composition-dependent entropic and energetic costs of conformations, as well as polymer-solvent interactions constants are just as important. The relationship between all these parameters is elusive and hard to be unambiguously identified, taken the number of relevant interactions present. Nevertheless, it is important to realize that a single parameter, such as $Q$,  allows to rationalize the salient IDPs' properties through their phase diagrams. For instance, one can speculate, that additional charge on IDPs and thus their increased persistence length is needed to compensate for the fact that they contain abnormal number of lightweight aminoacids and are thus very flexible. Thus it seems conceivable that by changing the counterion concentration in the cell it should be possible to tune also the IDP conformation.

\subsection*{Phase diagrams as important identifiers of proteins}

Considering the phase diagrams rather than particular conformations at particular environmental conditions, allows us to discern better the whole picture of protein behavior. This point of view is in fact especially important for such complex systems, as proteins undoubtedly are, and allows us to answer the question, posed at the beginning: since structurally IDPs belong to the same family of natural polypeptides as other proteins, they indeed behave similarly and do have the same shape of the phase diagram as the rest of the proteins. IDPs differ from other proteins only with respect to their position within this very same, universal protein folding phase diagram. 

We find it intriguing, that although the "skewed ellipsoid" phase diagram (Fig.~\ref{f0}) has been around for a while \cite{hawley} (for review, see, \emph{e.g.} \cite{smeller} and the whole volume of \emph{Biochem et Biophys Acta 1595 (2002)}), the idea that some proteins might gain order at higher temperatures or pressures, as indeed many IDPs do, is still not widely accepted. The question of description/classification of every particular protein would become much easier if the phase diagrams (both temperature-pressure and temperature-osmotic stress) for every protein would be shown together with its crystalline structure in the Protein Data Bank. This would be especially relevant for IDPs and the related re-considerations of the foundations of Protein Folding in general.

\subsection*{Limitations of our approach}

The major limitation of the GMPC model  is that it is intended to describe the helix-coil transition, not the folding \emph{per se}, and that since the detailed microscopic calculation of $Q$ is not within reach, it is rather a phenomenological model not straightforwardly related to microscopic parameters, as is also the case for other physical theories such as e.g. the theory of second order phase transitions or phenomenological liquid crystal theories. We are convinced that the GMPC model, while not directly describing the IDP folding transition, is instructive and its results enable a comparative analysis of secondary structure formation.
An important feature of our analysis is the introduction of the entropic cost for secondary structure formation, which allowed us to capture the peculiarities of the IDPs. While we have rationalized that higher values of $Q$ should hold for IDPs, we have to explicitly state, that we haven't calculated the real values of parameter $Q$, depending on primary structure. Rough estimates, made in Ref.~[\onlinecite{biopoly1}] resulted in the values $Q\approx64$ for polyglycine and $Q\approx36$ for polyalanine, within the range of values we have used. More detailed estimates require separate extensive study and cannot be done alongside the current paper. We plan to do these estimates in nearest future, especially after the important role of the entropic cost for hydrogen bond formation has been revealed.
In future we also plan to consider amyloid formation that strongly depends on H-bonding in $\beta$-sheets, another example of a secondary structure. The success in description we have achieved is mainly conditioned by the introduction of orientation-dependent water-polymer interactions and the phenomenological account of sequence-dependent entropic cost of secondary structure formation. Since the amyloid state formation depends on exactly the same factors, we expect interesting results from the future generalization of this approach.

\section*{Acknowledgements}

R.P. acknowledges support from the ARRS through Grants No. P1-0055 and J1-4297. 
Y.M. acknowledges support from the SCS RA through Grant No. 13-1F343 and from the NATO through Grant No. SfP-984537.

\section*{Author contributions}

A.B., Y.Sh.M, R.P. and V.A.P. contributed equally to the manuscript.

\section*{Competing financial interests}

The authors declare no competing interests, or other interests that might be perceived to influence the results and/or discussion reported in this paper.

\section*{Correspondence and requests}
Correspondence and requests for materials should be addressed to A.B.~(email: abadasyan@gmail.com).

\end{document}